\documentclass[aps,prx,reprint,showpacs,twocolumn,superscriptaddress,eqsecnum]{revtex4-2}
\usepackage{xr}
\usepackage{amsmath}
\usepackage{amsfonts,amssymb,amsthm,amsxtra}
\usepackage[ruled,vlined]{algorithm2e}
\usepackage[]{graphicx}
\usepackage{grffile}
\usepackage{mathrsfs}
\usepackage{framed}
\usepackage{bbm}
\usepackage{bm}
\usepackage{braket}
\usepackage{xcolor}
\usepackage{mathtools}
\usepackage[bookmarks=false,colorlinks=true,urlcolor=blue,citecolor=blue,linkcolor=blue]{hyperref}
\usepackage[]{qcircuit}

\DeclareMathOperator*{\argmin}{arg\,min}

\begin{document}

\title{Quantum dynamics simulations beyond the coherence time on NISQ hardware by variational Trotter compression}

\author{Noah F.~Berthusen}
\altaffiliation[Present address: ]{Department of Computer Science, University of Maryland, College Park, MD, 20742, USA}
\affiliation{Ames Laboratory, Ames, Iowa 50011, USA}
\affiliation{Department of Electrical and Computer Engineering, Iowa State University, Ames, Iowa 50011, USA}

\author{Tha\'{i}s V. Trevisan}
\affiliation{Ames Laboratory, Ames, Iowa 50011, USA}
\affiliation{Department of Physics and Astronomy, Iowa State University, Ames, Iowa 50011, USA}

\author{Thomas Iadecola}
\email{iadecola@iastate.edu}
\affiliation{Ames Laboratory, Ames, Iowa 50011, USA}
\affiliation{Department of Physics and Astronomy, Iowa State University, Ames, Iowa 50011, USA}

\author{Peter P.~Orth}
\email{porth@iastate.edu}
\affiliation{Ames Laboratory, Ames, Iowa 50011, USA}
\affiliation{Department of Physics and Astronomy, Iowa State University, Ames, Iowa 50011, USA}

\begin{abstract}
We demonstrate a post-quench dynamics simulation of a Heisenberg model on present-day IBM quantum hardware that extends beyond the coherence time of the device. This is achieved using a hybrid quantum-classical algorithm that propagates a state using Trotter evolution and then performs a classical optimization that
effectively compresses the time-evolved state into a variational form. When iterated, this procedure enables simulations to arbitrary times with an error controlled by the compression fidelity and a fixed Trotter step size. We show how to measure the required cost function, the overlap between the time-evolved and variational states, on present-day hardware, making use of several error mitigation methods. 
In addition to carrying out simulations on real hardware, we investigate the performance and scaling behavior of the algorithm with noiseless and noisy classical simulations. We find the main bottleneck in going to larger system sizes to be the difficulty of carrying out the optimization of the noisy cost function. 
\end{abstract}

\date{\today}

\maketitle

\section{Introduction}
Simulating quantum dynamics of interacting many-body systems is one of the main potential applications of quantum
computing, going back to Feynman's visionary work on simulating physics with quantum computers~\cite{
feynman_quantum_simulators}. For quantum many-body systems evolving under a generic nonintegrable Hamiltonian $H$,
such simulations are exponentially hard on classical devices due to the growth of entanglement during the simulation
time~\cite{Prosen_Znidaric-PRE-2007, schollwockDensitymatrixRenormalizationGroup2011}. Quantum processing units (QPUs), on the other hand, can efficiently simulate quantum dynamics with resources
scaling polynomially in the number of particles $N$ and in the simulation time $t$. Since Lloyd's original proposal~\cite{lloydUniversalQuantumSimulators1996}
to use a first-order Lie-Trotter product formula to decompose the unitary
time-evolution operator into elementary gates, several other methods have been discovered that exhibit a
more favorable asymptotic scaling of the required number of gates at or near the optimal scaling of $O(Nt)$~
\cite{Childs12,berryEfficientQuantumAlgorithms2007,berrySimulatingHamiltonianDynamics2015,Low17,childsFirstQuantumSimulation2018, childsNearlyOptimalLattice2019,lowWellconditionedMultiproductHamiltonian2019,haahQuantumAlgorithmSimulating2021}. Concrete resource estimates, however, show that this often comes at the cost of requiring a large number of ancilla qubits or having a large constant overhead of gates~\cite{childsFirstQuantumSimulation2018}. Therefore,
product formulas still remain a preferred choice on noisy intermediate-scale quantum (NISQ) hardware~\cite{
preskillQuantumComputingNISQ2018} due to their simplicity and competitive performance for physically relevant (e.g., local)
Hamiltonians that fulfill additional properties~\cite{Tran20}.

Despite this progress, direct quantum simulation algorithms 
face a critical drawback on NISQ QPUs:
the limited coherence time of the device imposes an upper bound on the depth of the quantum circuits that can be
implemented with high fidelity. This in turn upper-bounds the simulation time $t$ that can be reached before the output
is overwhelmed by errors---on current hardware, this time scale is roughly of order one in the natural units imposed
by the Hamiltonian being simulated~\cite{smithSimulatingQuantumManybody2019}. Quantum simulation beyond the coherence
time using ``fast-forwarding" algorithms~\cite{Cervera18,Cirstoiu20,gibbs2021longtime,YangBaxter} is possible, but only for nongeneric Hamiltonians, including those that can be mapped to free fermions~\cite{Atia17}.

Variational quantum algorithms (VQAs) provide a promising route to overcome 
the coherence-time obstacle for generic
Hamiltonians~\cite{cerezoVariationalQuantumAlgorithms2021,Bharti21NISQReview}. Their starting point is to represent the wavefunction by a variational ansatz $\ket{\psi(\boldsymbol{\vartheta})} = \prod_{j=1}^\mathcal{N} e^{-i \vartheta_j A_j} \ket{\psi_0}$ with $\mathcal{N}$ real parameters $\boldsymbol{\vartheta} = (\vartheta_1, \ldots, \vartheta_\mathcal{N})$, initial state $\ket{\psi_0}$, and Hermitian generators $A_j$, which are often chosen to be single Pauli strings or sums of commuting Pauli strings. To simulate quantum dynamics, the parameters $\vartheta_j$ are updated in a way that allows the variational state to follow the exact dynamics; different algorithms to perform the update have been proposed in the literature.

In one class of VQAs, one derives an equation of motion for the variational parameters, $\sum_{j} M_{ij} \dot{\vartheta}_j(t) = V_i$~\cite{Li_2017,Yuan19, Yao-AVQDS-PRX_Q-2021}, by extremizing the distance between the variational state and the exact time-evolved state at every (infinitesimally small) time step in the evolution.  
Here, the matrix $M_{ij}$ and vector $V_j$ 
must be obtained at every time step by performing measurements on the QPU, and
the main bottleneck of the algorithm is the large number of measurements. Since the number of components $M_{ij}$ scales quadratically with the number of variational parameters $\mathcal{N}$, the number of measurements $\mathcal{M}$ scales as $\mathcal{M} \propto \mathcal{N}^2$.

An alternative approach that reduces the number of measurements is to determine the time-dependence of the variational parameters $\boldsymbol{\vartheta}(t)$ by optimizing the state overlap fidelity $\mathcal{F} = |\braket{\psi(\boldsymbol{\vartheta})|\psi(t)}|^2$ between the variational state and the exact time-evolved state $\ket{\psi(t)}$~\cite{Lin_Pollmann-PRX_Q-2021,Barison_Carleo-Quantum-2021}.
In practice, the exact state can be approximated with high fidelity, for example, using Trotter evolution over a time interval $\tau$ sufficiently short to guarantee a desired high accuracy at finite gate depth.
Such a ``variational Trotter compression'' (VTC) approach combines the desirable aspects of Trotter evolution with the shallow gate depth requirements of variational methods.

Here, we expand on previous works that explored VTC algorithms on classical computers~\cite{Lin_Pollmann-PRX_Q-2021,Barison_Carleo-Quantum-2021} by performing the first implementation and benchmarking of a VTC algorithm on real quantum hardware. We explicitly demonstrate a simulation of quantum quench dynamics in a Heisenberg spin chain beyond the coherence time on the IBM Santiago QPU.
To achieve this goal, we employ several error mitigation strategies: we combine zero-noise extrapolation (ZNE) with Pauli twirling~\cite{temmeErrorMitigationShortDepth2017, Li_2017} and symmetry-based post-selection. At the current levels of noise on the QPU, we find it advantageous to avoid the gradient based optimization used in the variant of the VTC algorithm described in Ref.~\cite{Barison_Carleo-Quantum-2021}, and instead choose the non-gradient based genetic optimization algorithm CMA-ES~\cite{cma-es-code}.
We compare different ways to compute the overlap fidelity on QPUs, and find that a method based on a ``double time contour" circuit that foregoes the use of ancilla qubits and non-local SWAP gates~\cite{Buhrman_2001,schiffer2021adiabatic} is most suited for current NISQ hardware.
Finally, we demonstrate the scalability of the VTC algorithm both on noisy and noiseless quantum simulators up to system sizes of $M=6$ and $M=11$ sites, respectively.

The remainder of the paper is organized as follows: in Sec.~\ref{sec:algorithm} we present the general VTC algorithm and point out differences with previous works~\cite{Barison_Carleo-Quantum-2021,Lin_Pollmann-PRX_Q-2021}. We also discuss different quantum circuit implementations of the optimization cost function. Then, in Sec.~\ref{sec:application_Heisenberg_chain}, we apply VTC to simulate post quench dynamics in antiferromagnetic Heisenberg chains. In Sec.~\ref{subsec:Heisenberg_model_and_ansatz}, we introduce the model and describe the variational ansatz we use, and in Sec.~\ref{subsec:required_layer_number} we benchmark the capability of the ansatz to capture the exact time-evolved state. We then execute VTC on different classical simulators and on real quantum hardware. In Sec.~\ref{subsec:statevector}, we present results obtained on a statevector simulator. In Sec.~\ref{subsec:ideal_circuit_simulator}, we use an ideal circuit simulator to consider sample noise due to a finite number of quantum measurements, and in Sec.~\ref{subsec:noisy_circuit_simulator}, we show results from classical noisy circuit simulations, where the noise model parameters correspond to the IBM Santiago backend. Finally, in Sec.~\ref{subsec:qpu_santiago}, we demonstrate our key result: a quantum dynamics simulation beyond the qubit coherence time on the real IBM Santiago device. We conclude and discuss future research directions in Sec.~\ref{sec:conclusion}.

\section{Algorithm}
\label{sec:algorithm}
In this section we describe the different parts of the variational Trotter compression (VTC) algorithm.
\subsection{Choice of ansatz}
\label{subsec:ansatz_choice}
First, one needs to choose a particular way to build the variational ansatz $\ket{\psi(\boldsymbol{\vartheta})}$. The main part of the algorithm is independent of the form of the ansatz, but requires that it is able to faithfully represent the time-evolved state up to a desired fidelity. Since the entanglement content of the state increases during time evolution, the complexity of the ansatz and the number of required parameters grows with time $t$ (and also with system size $M$). We characterize this growth in detail for a specific model in Sec.~\ref{sec:application_Heisenberg_chain}. A key insight from previous works is that the growth with $t$ of the number of variational parameters can be much slower (linear versus exponential) than in purely classical algorithms based on matrix product states (MPSs)~\cite{Lin_Pollmann-PRX_Q-2021}, giving the VTC algorithm a potential quantum advantage.

The variational ansatz can either be chosen to have a fixed form or to be adaptively modified during the computation~\cite{grimsleyAdaptiveVariationalAlgorithm2019,qubit-ADAPT,Yao-AVQDS-PRX_Q-2021}. Here, we choose a fixed ansatz that is inspired by the Hamiltonian variational ansatz (HVA)~\cite{weckerProgressPracticalQuantum2015, Wiersema-PRX-2020} and that takes a layered form,
\begin{equation}
\ket{\psi(\boldsymbol{\vartheta})} = U(\boldsymbol{\vartheta})  \ket{\psi_0} = \prod_{l=1}^\ell \prod_{i=1}^N e^{-i \vartheta_{l,i} A_{i}} \ket{\psi_0}\,.
\end{equation}
Here, $\ell$ denotes the number of layers and $N$ is the number of variational parameters per layer. We choose the Hermitian operators $A_i$ to be sums of commuting Pauli strings that correspond to the terms in the Hamiltonian being simulated. Finally, $\ket{\psi_0}$ is an initial state that can be chosen arbitrarily; in particular, it does not need to correspond to the state of the system at the initial time $t_i$, which we denote by $\ket{\psi_i}$.

\subsection{Variational Trotter compression}
\label{subsec:compression_step}
The central part of the algorithm is the variational compression step. The key idea is to accurately propagate the variational state over a short, finite time interval $\tau$ using Trotter evolution, and then to re-express the resulting state, $U_\text{trot}(\tau) \ket{\psi(\boldsymbol{\vartheta}_{t})}$, in variational form via optimization of the overlap cost function
\begin{equation}
\label{eq:fidelity}
\mathcal{C} = |\braket{\psi(\boldsymbol{\vartheta}_{t+\tau})|U_\text{trot}(\tau)|\psi(\boldsymbol{\vartheta}_{t})}|^2 \,.
\end{equation}
The cost function is measured on the QPU and feeds into a classical optimization routine that determines the updated variational parameters as $\hat{\boldsymbol{\vartheta}}_{t+\tau} = \argmin_{\boldsymbol{\vartheta}_{t+\tau}} \mathcal{C}$. The complete compression step is then iterated until the final time $t_f$. 


The maximal value of $\tau$ in a single compression step is determined by the number of Trotter steps $n_{\text{max}}$ that can be executed on the noisy QPU, and by the desired Trotter error threshold $\varepsilon$, which is a function of $\tau/n_{\text{max}}$~\cite{lloydUniversalQuantumSimulators1996,Childs-PRX-2021}. There exist different methods to determine the overlap cost function on a QPU, which will be discussed in more detail below in Sec.~\ref{subsec:fidelity_implementation}. The simplest method concatenates the Trotter and the variational ansatz circuits and calculates the probability that the system returns to the initial state, \textit{i.e.},
\begin{equation}
\label{eq:fidelity_no_ancillas}
\mathcal{C} = |\braket{\psi_0| U^\dag(\boldsymbol{\vartheta}_{t+\tau}) U_{\text{trot}}(\tau) U(\boldsymbol{\vartheta}_{t}) | \psi_0}|^2 \,.
\end{equation}
Here, $U_{\text{trot}}(\tau)$ is a (first-order) Trotter circuit using $n$ steps. We find this straightforward method to be robust under noise and preferable on NISQ hardware with nearest-neighbor connectivity compared to alternative methods that use ancilla qubits or multiple SWAP operations~\cite{Buhrman_2001,schiffer2021adiabatic}. Since the variational and Trotter circuits are applied consecutively in Eq.~\eqref{eq:fidelity_no_ancillas}, the depth of the ansatz, which can be expressed as the number of layers $\ell$, must be taken into account when choosing the number of Trotter steps $n$. Loosely speaking, one must have that $2 \ell + n < n_\text{max}$, where $n_\text{max}$ is the maximal number of Trotter steps that can be executed within the finite coherence time of the device.

Different methods can be used for the classical optimization of the overlap cost function $\mathcal C$. Ref.~\cite{Barison_Carleo-Quantum-2021} proposes a gradient-based optimizer, where the gradient is measured directly using a quantum circuit, while Ref.~\cite{Lin_Pollmann-PRX_Q-2021} performs the optimization purely classically using tensor network based methods. We have compared the performance of gradient- and non-gradient-based methods and find that non-gradient-based methods such as CMA-ES~\cite{cma-es-code} are preferable for noisy cost functions when considering realistic noise levels present on current hardware.

\subsection{Quantum circuit implementation of cost function measurement}
\label{subsec:fidelity_implementation}
As mentioned previously, when implementing the VTC algorithm on a real quantum device one needs to decide how to evaluate the overlap cost function $\mathcal{C}$ in Eq.~\eqref{eq:fidelity}, since the full wavefunction is not accessible. We consider two methods to evaluate $\mathcal C$ that are physically realizable on current NISQ devices, and we compare their resource scaling. Since the fidelities of two-qubit entangling gates are much lower than those of single-qubit gates on current hardware, we focus on the scaling of the number of two-qubit gates.

\subsubsection{SWAP test circuit}
\begin{figure}
    \centering
    \begin{equation*}
    \Qcircuit @C=1.5em @R=1em @!R {\lstick{\ket{0}}& \gate{H} & \ctrl{1} & \gate{H} & \qw & \meter \\
    \lstick{\ket{\Phi}} & \qw & \qswap & \qw & \qw & \qw\\
    \lstick{\ket{\Psi}} & \qw & \qswap \qwx & \qw & \qw & \qw 
    }
\end{equation*}
    \caption{SWAP-test circuit diagram for computing the overlap between two quantum states $\ket{\Psi}$ and $\ket{\Phi}$ via the expectation value $\langle Z \rangle = |\braket{\Psi | \Phi}|^2$ of an ancilla qubit. }
    \label{fig:swap_test}
\end{figure}
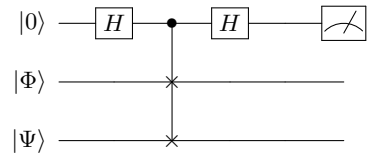

One common method for computing the overlap of two quantum states is the SWAP test~\cite{Buhrman_2001,fouldsControlledSWAPTest2021}, whose
circuit implementation is shown in Fig.~\ref{fig:swap_test}. Here, the two crosses denote swap operations between pairs of individual qubits in the two quantum registers containing the states $\ket{\Phi}$ and $\ket{\Psi}$. The ancilla expectation value $\langle Z \rangle = |\braket{\Psi | \Phi}|^2$ is equal to the overlap of the states and thus serves 
as a cost function for VTC. Specifically, we initialize the states $\ket{\Phi} = U_{\text{trot}}(\tau) U(\boldsymbol{\vartheta}_t)\ket{\psi_0}$ and $\ket{\Psi} = U(\boldsymbol{\vartheta}_{t+\tau})\ket{\psi_0}$ and find the optimal parameters $\bm\vartheta_{t+\tau}$ by maximizing $\langle Z \rangle$ for the ancilla.

Let us now estimate the number of two-qubit gates required to implement this circuit on two quantum registers of length $M$, each corresponding to a system of $M$ qubits.
Swapping arbitrary $M$-qubit states $\ket{\Phi}$ and $\ket{\Psi}$ requires $M$ controlled-SWAP gate operations, each of which can be broken down into two controlled-NOT (CNOT) gates and one Toffoli gate. The Toffoli gate
can be further decomposed into a circuit containing 6 CNOTs. Therefore, the swap test incurs a cost of $8M$ CNOTs in total.
One major drawback of using the SWAP test on QPUs with local qubit connectivity, however, is that the ancilla must be coupled to every other qubit during the cascading controlled-SWAPs. As the current generation of superconducting QPUs lacks such nonlocal connectivities, one would need to compile a nonlocal SWAP into many local CNOTs, which significantly increases the resource cost of the SWAP test.

\subsubsection{Double time contour circuit}
\label{subsubsec:loschmidt_echo_circuit}
\begin{figure}
    \centering
    \begin{equation*}
    \Qcircuit@C=1.3em @R=0.6em {
    && \multigate{4}{U(\boldsymbol{\vartheta}_{t})} & \multigate{4}{U_{\text{trot}}(\tau)} & \multigate{4}{U^\dag(\boldsymbol{\vartheta}_{t+\tau})} & \meter \\
    && \ghost{U(\boldsymbol{\vartheta}_{t})} & \ghost{U_{\text{trot}}(\tau)} & \ghost{U^\dag(\boldsymbol{\vartheta}_{t+\tau})} & \meter \\
    && \ghost{U(\boldsymbol{\vartheta}_{t})} & \ghost{U_{\text{trot}}(\tau)} & \ghost{U^\dag(\boldsymbol{\vartheta}_{t+\tau})} & \meter\\
    && \ghost{U(\boldsymbol{\vartheta}_{t})} & \ghost{U_{\text{trot}}(\tau)} & \ghost{U^\dag(\boldsymbol{\vartheta}_{t+\tau})} & \meter\\
    && \ghost{U(\boldsymbol{\vartheta}_{t})} & \ghost{U_{\text{trot}}(\tau)} & \ghost{U^\dag(\boldsymbol{\vartheta}_{t+\tau})} & \meter \\
    {\inputgroupv{1}{5}{0.5em}{4em}{\ket{\psi_0}}}
    }
\end{equation*}
    \caption{Double time contour quantum circuit for ancilla-free evaluation of the overlap cost function $\mathcal{C}$ in Eq.~\eqref{eq:fidelity_no_ancillas}. Circuit is shown for $M=5$. }
    \label{fig:loschmidt_echo}
\end{figure}
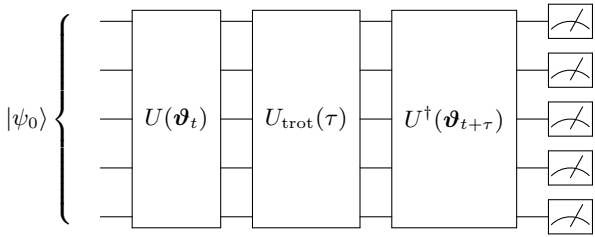

The simplest way to evaluate the overlap cost function, which does not require any qubit overhead or ancillas, is to implement a ``double time contour'' circuit, corresponding to the direct implementation of the overlap matrix element in Fig.~\ref{fig:loschmidt_echo}. The key idea is that the inverse of the updated variational circuit $U^\dag(\boldsymbol{\vartheta}_{t+\tau})$ effectively unwinds the evolution induced by $U_{\text{trot}}(\tau) U(\boldsymbol{\vartheta}_t)$. The probability $p_{\ket{\psi_0}}$ for the system to end up in its initial state $\ket{\psi_0}$ is maximal for the optimal parameters $\hat{\boldsymbol{\vartheta}}_{t+\tau}$. We find this cost function to be robust to noise under realistic conditions. The circuit is shown in Fig.~\ref{fig:loschmidt_echo} and consists of a consecutive application of three unitary circuits onto a fixed initial state $\ket{\psi_0}$. For convenience, we choose the initial state $\ket{\psi_0}$ to be a $Z$-basis state.
This method to evaluate $\mathcal C$ does not require any additional overhead and the number of required qubits is equal to $M$, the number of qubits in the system. 
Note, however, that the depth of the circuit $U^\dag(\boldsymbol{\vartheta}_{t+\tau})U_{\text{trot}}(\tau) U(\boldsymbol{\vartheta}_t)$ is about $1.5$ times larger than that required for the SWAP test.

\section{Application to Heisenberg spin chain dynamics}
\label{sec:application_Heisenberg_chain}
In this section, we apply the variational Trotter compression algorithm to investigate post-quench dynamics in antiferromagnetic Heisenberg chains. We first describe the variational ansatz we employ and then present results using a statevector simulator, ideal and noisy circuit simulators, and real quantum hardware.
\subsection{Model Hamiltonians and variational ansatz}
\label{subsec:Heisenberg_model_and_ansatz}
To benchmark the VTC algorithm and to study its robustness with respect to noise, we apply it to investigate quantum quench dynamics in integrable and nonintegrable Heisenberg spin chains. The Hamiltonian of the pure Heisenberg chain, which is integrable, is given by
\begin{equation}
    H_0 = \frac{J}{4}\sum^M_{i=1} \left( X_iX_{i+1} + Y_iY_{i+1} + Z_iZ_{i+1} \right)\,.
    \label{eq:Hnn}
\end{equation}
Here, $X_i,Y_i,Z_i$ are Pauli operators at site $i$, and $M$ denotes the total number of sites of the chain. For concreteness, we will focus in the following on the antiferromagnetic model with $J > 0$, and consider the quantum spin dynamics that arises when initially preparing the system in the classical N\'eel ground state $\ket{\psi_i} = \ket{0101\cdots}$. The classical N\'eel state is not an eigenstate of $H_0$, leading to nontrivial dynamics of $\ket{\psi(t)} = e^{-iH_0 t} \ket{\psi_i}$.
To study the difference between simulations of integrable versus nonintegrable dynamics, we also consider an integrability breaking next-neighbor interaction in the model,
\begin{equation}
    H_1 = H_0 + \frac{J}{4}\sum^M_{i=1} Z_iZ_{i+2} \,.
    \label{eq:Hnnn}
\end{equation}
Unless explicitly stated otherwise, we employ periodic boundary conditions in the following.

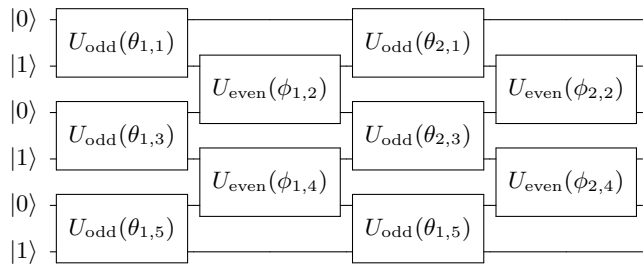
\begin{figure}[tbh]
\centering
\begin{equation*}
    \Qcircuit @C=0.25em @R=1.em {
    \lstick{\ket{0}} & \multigate{1}{U_{\text{odd}}(\theta_{1,1})} & \qw & \qw & \qw & \multigate{1}{U_{\text{odd}}(\theta_{2,1})} & \qw & \qw & \qw
    \\
    \lstick{\ket{1}} & \ghost{U_{\text{odd}}(\theta_{1,1})} & \qw & \multigate{1}{U_{\text{even}}(\phi_{1,2})} & \qw & \ghost{U_{\text{odd}}(\theta_{2,1})} & \qw & \multigate{1}{U_{\text{even}}(\phi_{2,2})} & \qw
    \\
    \lstick{\ket{0}} & \multigate{1}{U_{\text{odd}}(\theta_{1,3})} & \qw & \ghost{U_{\text{even}}(\phi_{1,2})} & \qw & \multigate{1}{U_{\text{odd}}(\theta_{2,3})} & \qw & \ghost{U_{\text{even}}(\phi_{2,2})} & \qw
    \\
    \lstick{\ket{1}} & \ghost{U_{\text{odd}}(\theta_{1,3})} & \qw & \multigate{1}{U_{\text{even}}(\phi_{1,4})} & \qw & \ghost{U_{\text{odd}}(\theta_{2,3})} & \qw & \multigate{1}{U_{\text{even}}(\phi_{2,4})} & \qw
    \\
    \lstick{\ket{0}} & \multigate{1}{U_{\text{odd}}(\theta_{1,5})} & \qw & \ghost{U_{\text{even}}(\phi_{1,4})} & \qw & \multigate{1}{U_{\text{odd}}(\theta_{1,5})} & \qw & \ghost{U_{\text{even}}(\phi_{1,4})} & \qw
    \\
    \lstick{\ket{1}} & \ghost{U_{\text{odd}}(\theta_{1,5})} & \qw & \qw & \qw & \ghost{U_{\text{odd}}(\theta_{1,5})} & \qw & \qw & \qw
    }
\end{equation*}
\caption{Quantum circuit implementing the brickwall ansatz $\ket{\psi(\bm{\vartheta}^{(\ell)})}$ in Eq.~\eqref{eq:ansatz} for $\ell = 2$ and $M=6$. For simplicity, we show the circuit for OBC. For PBC, there exists an additional even layer gate $U_{\text{even}}(\phi_{l,M})$ between the first and the last qubit. 
}
\label{fig:brickwall_circuit}
\end{figure}

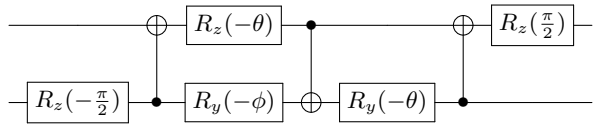
\begin{figure}[tbh]
\centering
\begin{equation*}
    \label{eq:n_circuit}
    \Qcircuit @C=.75em @R=1.5em {
    & \qw & \targ & \gate{R_z(-\theta)} & \ctrl{1} & \qw & \targ & \gate{R_z(\frac{\pi}{2})} & \qw
    \\
    & \gate{R_z(-\frac{\pi}{2})} & \ctrl{-1} & \gate{R_y(-\phi)} & \targ & \gate{R_y(-\theta)} & \ctrl{-1} & \qw & \qw
    }
\end{equation*}
\caption{Quantum circuit implementing the unitary operator $U(\alpha) = \exp\bigl[ -i \alpha (X_j X_{j+1} + Y_j Y_{j+1} + Z_j Z_{j+1})\bigr]$, where $\theta = \frac{\pi}{2} - 2\alpha$ and $\phi = 2\alpha - \frac{\pi}{2}$. }
\label{fig:Heisenberg_circuit}
\end{figure}

Let us now discuss our choice of variational ansatz. For the integrable case, we consider a brickwall-type quantum circuit with $\ell$ layers (shown in Fig.~\ref{fig:brickwall_circuit}) that we apply to an initial product state $\ket{\psi_0} = \ket{0101\cdots}$:
\begin{equation}
    \label{eq:ansatz}
    \ket{\psi(\bm\vartheta^{(\ell)})} = \prod^\ell_{l=1} U_{\rm even}(\bm\phi_l) \, U_{\rm odd}(\bm\theta_l) \ket{\psi_0}\,.
\end{equation}
Here, $\bm\vartheta^{(\ell)} \equiv (\bm\theta_1,\bm\phi_1,\dots,\bm\theta_\ell,\bm\phi_\ell)$ are $M \ell$ variational parameters and
\begin{subequations}
\begin{align}
    U_{\rm odd}(\bm\theta_l)&=\prod_{j\, \text{odd}}e^{-i\, \theta_{l,j}\, (X_jX_{j+1}+Y_jY_{j+1}+Z_jZ_{j+1})}\\
    U_{\rm even}(\bm\phi_l)&=\prod_{j\, \text{even}}e^{-i\, \phi_{l,j}\, (X_jX_{j+1}+Y_jY_{j+1}+Z_jZ_{j+1})}
\end{align}
\label{eq:U_even_odd_ansatz}
\end{subequations}
are the unitary operators acting on the odd and the even bonds of the chain, respectively. Here, $\theta_{l,j}$ is the $j$th entry of the parameter vector $\bm \theta_l$, and $\phi_{l,j}$ is the $j$th entry of $\bm \phi_l$. The quantum circuit has depth $2 \ell$, corresponding to the number of unitaries applied to every qubit. Note that the length of the vector $(\bm\theta_l,\bm\phi_l)$ is equal to the number of sites $M$ and $\bm\vartheta^{(\ell)}$ has $M \ell$ components. 
Additionally, for odd $M$, the ansatz contains a boundary term $U_{\text{boundary}}$ that does not commute with either $U_{\rm odd}$ or $U_{\rm even}$. A compact quantum circuit representation of the unitary operators in Eqs.~\eqref{eq:U_even_odd_ansatz} has been given in Ref.~\onlinecite{Vatan04} and is shown in Fig.~\ref{fig:Heisenberg_circuit}.

For the nonintegrable model $H_1$, we add an additional unitary to each brickwall layer $l$:
\begin{align}
    U_{Z} (\bm \gamma_l) = \prod_{j=1}^M e^{-i\, \gamma_{l, j}\,Z_j Z_{j+2}}\,.
\end{align}
This doubles the number of parameters in the ansatz, which is given by $2 M \ell$ in the nonintegrable case.

\subsection{Required number of layers $\ell$}
\label{subsec:required_layer_number}
To benchmark the ability of the variational ansatz in Eq.~\eqref{eq:ansatz} to represent the exact wavefunction within a desired accuracy $\epsilon$, we
determine the minimal number of layers $\ell_\text{min}(t, \epsilon, M)$ needed such that the variational ansatz can represent the exact time-evolved state $\ket{\psi(t)}$ at time $t$ up to an infidelity $1 - \mathcal{F} < \epsilon$. To do this, for each $t$ we numerically minimize the infidelity
\begin{equation}
    \label{eq:exact_cf}
    1 - \mathcal{F}\bigl(t, \bm \vartheta^{(\ell)} \bigr) = 1- |\braket{\psi(\bm\vartheta^{(\ell)})|\psi(t)}|^2
\end{equation}
over the parameters $\boldsymbol{\vartheta}^{(\ell)}$ to obtain the optimal parameters $\hat{\bm \vartheta}^{(\ell)}$. Repeating this for different values of $\ell$, we find $\ell_\text{min}(t, \epsilon, M)$ as the smallest $\ell$ for which the minimal infidelity $1-\mathcal{F}(t, \hat{\bm \vartheta}^{(\ell)})$ falls below the desired error threshold $\epsilon$. Here, we obtain the exact state $\ket{\psi(t)} = e^{-iH t} \ket{\psi_i}$ via exact diagonalization with $H = H_0$ (or $H = H_1$) and initial state $\ket{\psi_i} = \ket{0101\cdots}$. The optimization is performed using the gradient-based optimizer L-BFGS-B~\cite{florian_gerber_2020_3888570}.
\begin{figure}
    \centering
    \includegraphics[width=\linewidth]{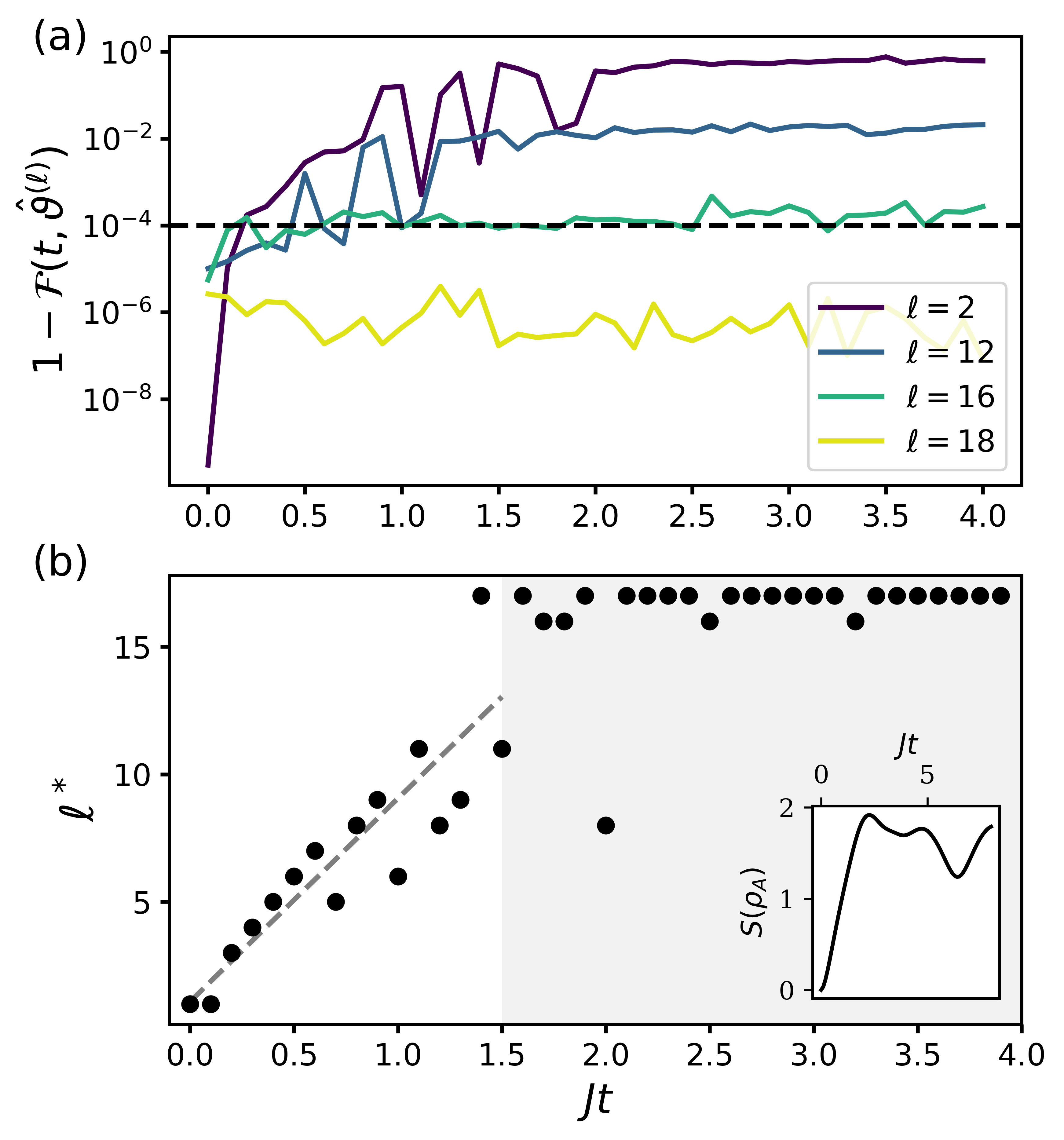}
    \caption{(a) Minimal infidelity $1 - \mathcal{F}(t, \hat{\bm \vartheta}^{(\ell)})$ as a function of time $t$ for different values of $\ell$. Results are for the $M=8$ pure Heisenberg model $H_0$, Eq.~\eqref{eq:Hnn}, using PBC. We obtain the infidelity by performing a numerical optimization of Eq.~\eqref{eq:exact_cf} every $0.1 Jt$. (b) Required layer number $\ell^*(t) \equiv \ell_{\rm min}(t, \epsilon=1 \times 10^{-4}, M=8)$ as a function of time. The inset shows the dynamics of the half-chain entanglement entropy for comparison.}
    \label{fig:t_vs_fidelity}
\end{figure}

We first study the time-dependence of the minimal infidelity for a fixed number of layers $\ell$.  As shown in Fig.~\ref{fig:t_vs_fidelity}, $1-\mathcal{F}(t, \hat{\bm \vartheta}^{(\ell)})$ first increases rapidly after the quench (unless $\ell$ is sufficiently large) and then saturates at a value that decreases with increasing layer number. 
This can be understood by comparison to the behavior of the entanglement entropy when tracing out half the system's degrees of freedom [see inset of Fig.~\ref{fig:t_vs_fidelity}(b)]: it grows linearly over time until it reaches saturation due to the finite system size. Since each layer of the brickwall quantum circuit couples only nearest-neighbor qubits, the spread of entanglement across the system in the variational state is limited by the total number of layers $\ell$. As a result, for a fixed layer number $\ell$, the infidelity $1 - \mathcal{F}$ grows as a function of time as the variational ansatz is unable to capture the entanglement that builds up in the system during time evolution.

In the bottom panel of Fig.~\ref{fig:t_vs_fidelity}, we show the growth with time of the layer number $\ell^*(t)$ required to keep the infidelity at time $t$ below a fixed threshold that we set to $\epsilon = 1 \times 10^{-4}$ (see also dashed line in the top panel). Like the entanglement entropy, the required layer number grows linearly in time before it reaches saturation. This linear growth in the number of layers and variational parameters provides an opportunity for quantum advantage, since one generically expects state-of-the-art classical techniques based on matrix product techniques to exhibit an exponential scaling with time of the number of parameters~\cite{Prosen_Znidaric-PRE-2007,Lin_Pollmann-PRX_Q-2021}.
Finally, due to finite system size, the required layer number saturates at long times. We denote the saturation value of the layer number by $\ell^* \equiv \lim_{t \rightarrow \infty} \ell^*(t, \epsilon, M)$. In practice, we choose a time $t_f$ much larger than the saturation timescale and define $\ell^* \equiv \ell^*(t_f)$. Note that an ansatz with $\ell^*$ layers is able to represent the time-evolved state out to arbitrary times (see also top panel of Fig.~\ref{fig:t_vs_fidelity}).

\begin{figure}
    \centering
    \includegraphics[width=\linewidth]{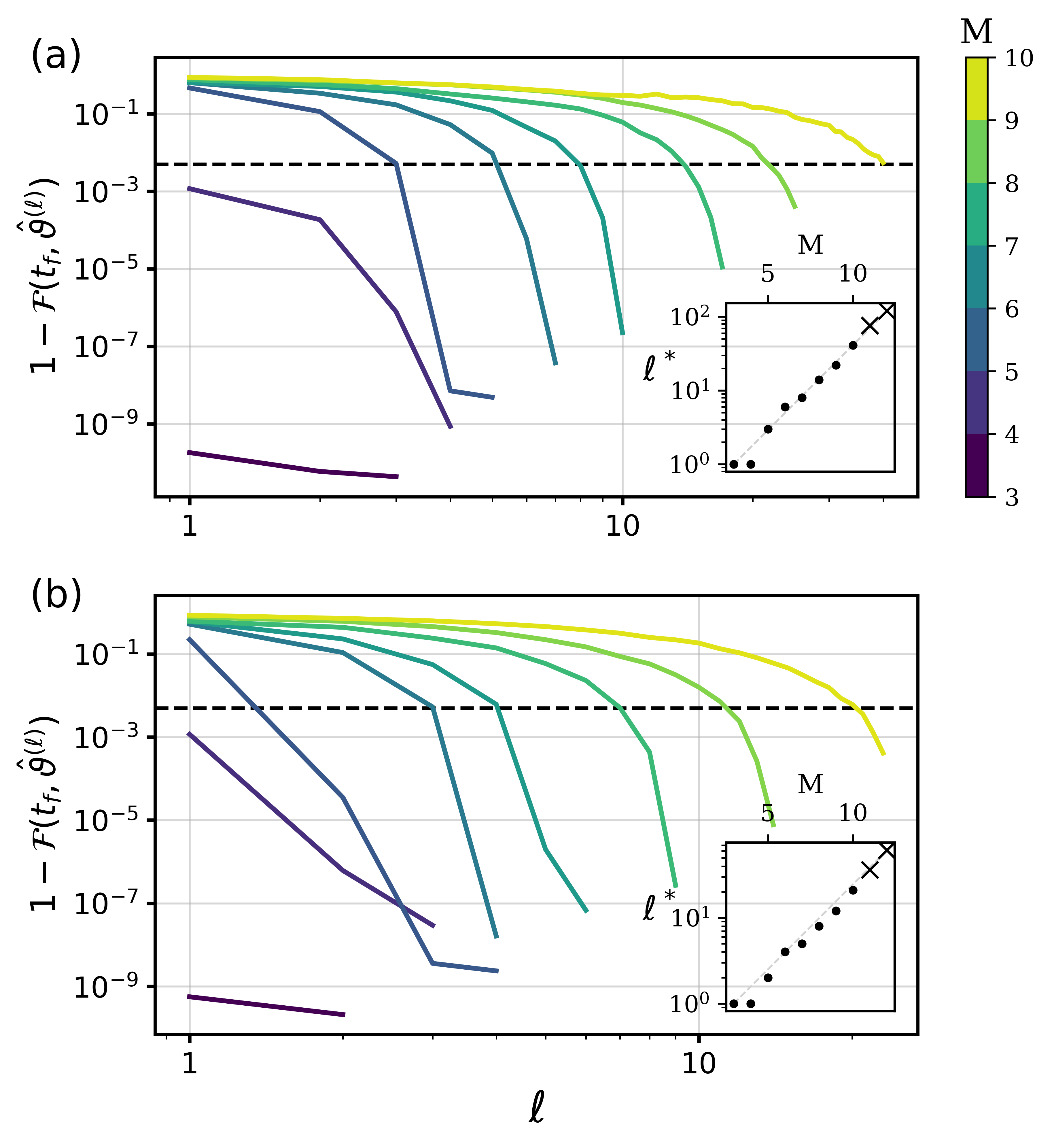}
    \caption{Minimal infidelity $1-\mathcal{F}(t_f, \hat{\bm \vartheta}^{(\ell)})$ at late time $t_f = 50 J^{-1}$ as a function of layer number $\ell$. Different curves are for different system sizes $M$. Panel (a) is for the pure Heisenberg model $H_0$, Eq.~\eqref{eq:Hnn}, and panel (b) is for the nonintegrable model $H_1$, Eq.~\eqref{eq:Hnnn}. Both results are for PBC. The black dashed line shows a fixed error threshold $\epsilon = 5 \times 10^{-3}$, which defines $\ell^*$. The insets show that $\ell^*$ exhibits exponential scaling with system size $M$. Since we find that $1-\mathcal{F}$ decays exponentially with increasing $\ell$, the values of $\ell^*$ denoted by a cross are obtained by exponential extrapolation of $1-\mathcal{F}(\ell)$.
    }
    \label{fig:p_star}
\end{figure}

We now systematically study the dependence of the saturation layer number $\ell^*$ on system size $M$ and desired error threshold $\epsilon$. We therefore fix a final time $t_f = 50 J^{-1}$, which is larger than the saturation time scale of the entanglement entropy for all system sizes we consider.
In Fig.~\ref{fig:p_star} we show the minimal infidelity, $1-\mathcal F(t_f, \hat{\bm \vartheta}^{(\ell)})$, as a function of layer number $\ell$. We present results for the integrable Heisenberg model $H_0$ in panel (a) and for the nonintegrable model $H_1$ in panel (b). As a function of $\ell$, the infidelity curves are first relatively flat until a characteristic $M$-dependent value, where they start to plunge to much smaller values. For definiteness, we set the desired error threshold to $\epsilon = 5 \times 10^{-3}$, which is indicated by the black dashed line in Fig.~\ref{fig:p_star}. The insets show the resulting $\ell^*(\epsilon = 5 \times 10^{-3}, M)$ as a function of system size $M$. We find that $\ell^*$ grows exponentially with $M$, which implies an exponential growth of the number of variational parameters $\mathcal{N} \propto M \ell$ that are needed to describe the long time dynamics. For $M=10$ we find $\ell^*(M=10) \approx 40$ in the integrable model and $\ell^*(M=10) \approx 20$ in the nonintegrable one. In both cases, the number of variational parameters $\mathcal{N} \approx 350$, since the nonintegrable ansatz has twice as many parameters per layer as the integrable one. This exponential scaling behavior at large times constitutes a bottleneck of the approach when considering larger system sizes.  Note, however, that at short times we find a favorable linear scaling of $\ell$ and $\mathcal{N}$ with time $t$, which provides an opportunity for quantum advantage. Such behavior was also reported previously using a sequential quantum circuit ansatz~\cite{Lin_Pollmann-PRX_Q-2021}.

\subsection{Statevector simulator results}
\label{subsec:statevector}
We now discuss the performance of the VTC algorithm in simulating post-quench dynamics of the antiferromagnetic Heisenberg chain. In the following we focus on the integrable model $H_0$, since the results of Fig.~\ref{fig:p_star} indicate that the algorithm's performance will not differ substantially between the two cases. In this section, we discuss results obtained using an exact statevector simulator, where we have direct access to the overlap cost function $\mathcal C$ in Eq.~\eqref{eq:fidelity}. 
Recall that the VTC algorithm consists of two steps: a propagation step, where we use a first-order Trotter product formula to evolve the state from time $t$ to $t + \tau$, and a compression step, where the time-evolved state is compressed into variational form by numerical optimization of the overlap cost function $\mathcal C$.

\begin{figure}[tb]
    \centering
    \includegraphics[width=\linewidth]{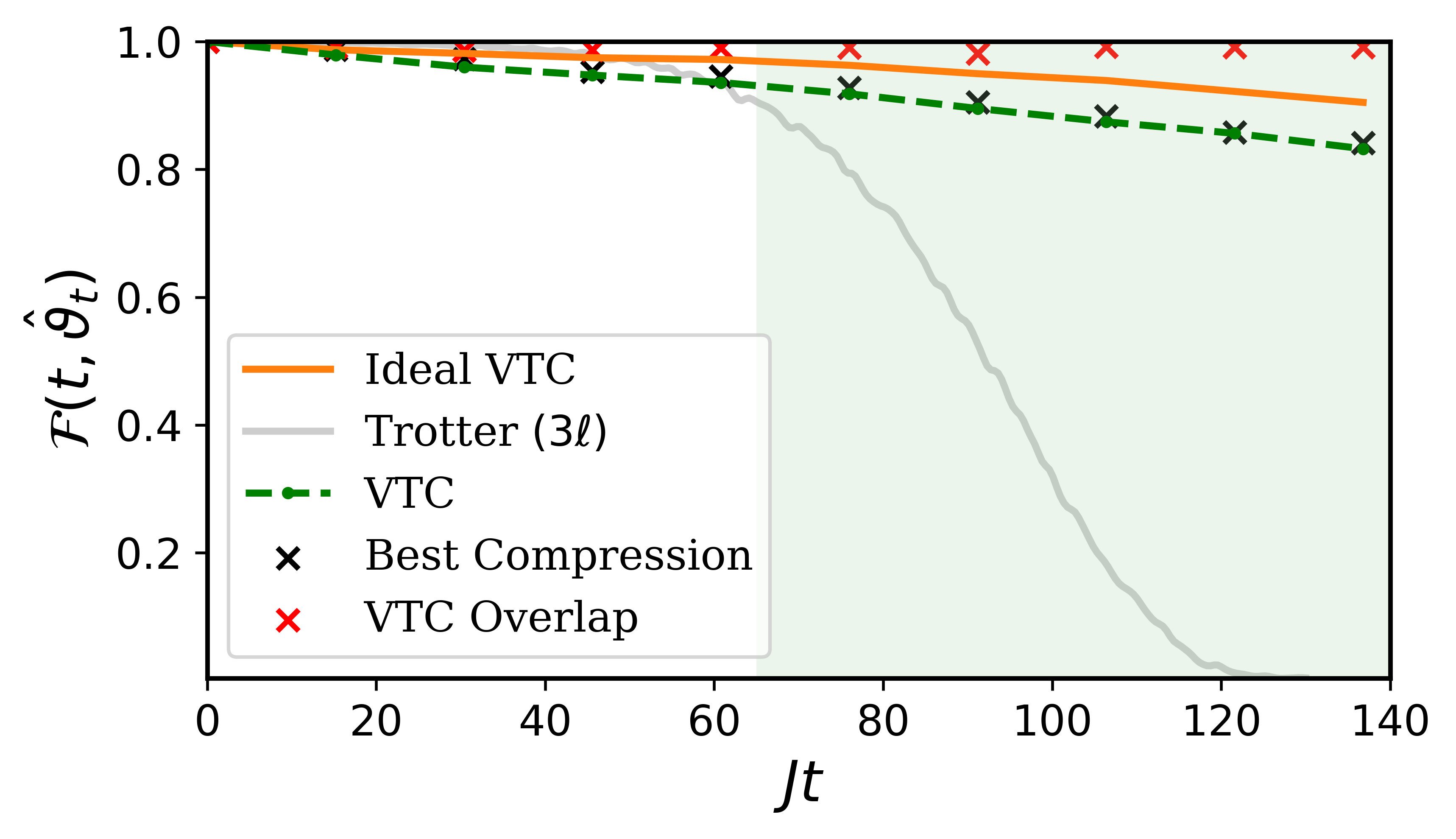}
    \caption{Statevector simulator results for the VTC state overlap fidelity $\mathcal{F} = |\braket{\psi(\hat{\bm \vartheta}_t)| \psi(t)}|^2$ (green dashed) with the exact state $\ket{\psi(t)}$ during post-quench dynamics in the $M=11$ pure Heisenberg chain $H_0$, Eq.~\eqref{eq:Hnn}. The system is initially prepared in the classical N\'eel state $\ket{\psi_i}$. We use $\ell = 76$ layers in the variational ansatz, and $n = 76$ Trotter steps in the VTC propagation step from $t \rightarrow t + \tau$ with $\tau = 15.2$. The error threshold during compression is set to $\epsilon = 5 \times 10^{-3}$ and we employ the gradient based optimizer L-BFGS-B. The orange curve depicts the Trotter simulation fidelity $\mathcal F_{\rm Trot}$ with fixed step size $\tau/n = 0.2$, corresponding to VTC with perfect compression.
    The grey curve are results from direct Trotter simulations, where we use a fixed number of $3 \times 76 = 228$ Trotter steps for a fair comparison. In the region highlighted in green, $t > 60 J^{-1}$, the state fidelity is larger for VTC than for direct Trotter simulation, showing an advantage at larger times. The black crosses denote the fidelity at time $t$ of the best possible Trotter compression state $U_{\text{Trot}} \ket{\psi(\hat{\bm \vartheta}_{t-\tau})}$ , i.e., $|\braket{\psi(t) | U_{\text{Trot}} | \psi(\hat{\bm \vartheta}_{t-\tau})}|^2$. Red crosses denote the overlap between the best possible compression state and the state to which the algorithm converged during the compression optimization step, i.e., $|\braket{ \psi(\hat{\bm \vartheta}_{t}) | U_{\text{Trot}} | \psi(\hat{\bm \vartheta}_{t-\tau})}|^2$, which is equal to $\mathcal{C}$ at the end of the optimization.
    }
    \label{fig:statevector_M_11}
\end{figure}

In Fig.~\ref{fig:statevector_M_11}, we present results for the state overlap fidelity of the VTC-evolved state with the exact state: $\mathcal{F}(t) = |\braket{\psi(\hat{\bm \vartheta}_t)| \psi(t)}|^2$, where $\ket{\psi(t)} = e^{-iH_0 t} \ket{\psi_i}$ and $\hat{\bm \vartheta}_t$ are the optimal variational parameters maximizing the cost function $\mathcal C$ at time $t$. Note that in this and the following sections, we drop the $(\ell)$ superscript when referring to the variational parameters. Instead, we always use $\ell^* \equiv \lim_{t \rightarrow \infty} \ell^*(t, \epsilon, M)$ for a specified $\epsilon$ and $M$.  The VTC fidelity $\mathcal{F}(t)$ is shown as a green dashed line in Fig.~\ref{fig:statevector_M_11}. Here, we focus on simulations at system size $M=11$ over a long time $t_f = 140 J^{-1}$ using a variational ansatz with $\ell = 76$ layers that is able to represent the exact time-evolved state over the full time interval. We find a sizable fidelity $\mathcal{F}(t_f) = 0.83$ at the end of the simulation after performing 9 propagation and compression steps at times $t = m \tau$ with $\tau = 15.2$ and $m = 1, \ldots, 9$. Here, we have chosen $n = \ell$ Trotter steps to propagate the state from time $t$ to $t + \tau$.

This value of $\mathcal{F}(t_f)$ should be contrasted with the fidelity obtained by direct Trotter simulation using the same resources as VTC: the Trotter simulation fidelity $\mathcal{F}_{\text{Trot}}(t) = |\braket{\psi_\text{Trot}(t)| \psi(t)}|^2$ falls to zero already at time $t = 120 J^{-1}$ (grey curve). We note that the direct Trotter simulation uses a fixed number of $3 \ell$ Trotter steps. This is because the double time contour circuit used to measure the VTC cost function $\mathcal C$ is of length $3 \ell$ if one chooses $n=\ell$ during the VTC propagation step~[see Eq.~\eqref{eq:fidelity_no_ancillas}].

During the VTC algorithm there are two independent sources of error that occur during the propagation and the compression step, respectively. First, a Trotter error occurs during state propagation that is controlled by the Trotter step size $\tau/n$. Second, a compression error arises that is given by the final (i.e., minimal) value of $1-\mathcal C$ at the end of the numerical optimization. This error is controlled by the optimization parameter $\epsilon$ that sets the threshold for convergence. Of course, a smaller value of $\epsilon$ makes the optimization more difficult and time consuming. To contrast these two sources of error, we include in Fig.~\ref{fig:statevector_M_11} results for a perfect compression (orange line), corresponding to the Trotter simulation fidelity $\mathcal F_{\rm Trot}(t)$ with fixed step size, where the only source of error is the finite, fixed Trotter step size $\tau/n$. In the ideal case of perfect compression, this error could be brought down arbitrarily by reducing $\tau$. However, since we have to perform more compression steps for smaller $\tau$, this will increase the compression error in practice. This leads to our choice of parameters for the VTC simulation in Fig.~\ref{fig:statevector_M_11}, where the two errors are comparable.

Finally, the main bottleneck of the algorithm in going to larger system sizes is the time needed to perform the compression optimization. Due to the large number of parameters, e.g. $\mathcal{N} = M \ell = 836$ for $M=11$ and $\ell = 76$, the optimization of the variational ansatz to produce a single compression point in Fig.~\ref{fig:statevector_M_11} takes a few hours of CPU time, preventing long-time simulations for larger systems. On the other hand, if one limits the simulation to shorter times, once can reduce the number of layers $\ell$ in the variational ansatz while keeping the product $\mathcal{N} = M \ell$ constant, and thereby simulate larger systems. This was demonstrated for a mixed-field Ising model using a different classical optimization method based on tensor network techniques in Ref.~\cite{Lin_Pollmann-PRX_Q-2021}.

\subsection{Ideal quantum circuit simulator results}
\label{subsec:ideal_circuit_simulator}
Next, we investigate the performance of the VTC algorithm on a noiseless quantum circuit simulator, where we evaluate the overlap cost function using either the SWAP-test (Fig.~\ref{fig:swap_test}) or the double time contour circuit (Fig.~\ref{fig:loschmidt_echo}) using a finite number of quantum measurements, or ``samples." The fact that the overlap cost function now exhibits sample noise increases the difficulty of the numerical optimization during the compression step. In particular, we find that the non-gradient-based optimizer CMA-ES is more reliable and converges much faster to a minimum of the noisy overlap cost function compared to a gradient-based optimizer such as L-BFGS-B. While the sample noise can in principle be brought arbitrarily low by increasing the number of samples, we here focus on a realistic number of samples, between $2^{14}$ and $2^{16}$, that could be executed on current NISQ hardware.

\begin{figure}[t]
    \centering
    \includegraphics[width=\linewidth]{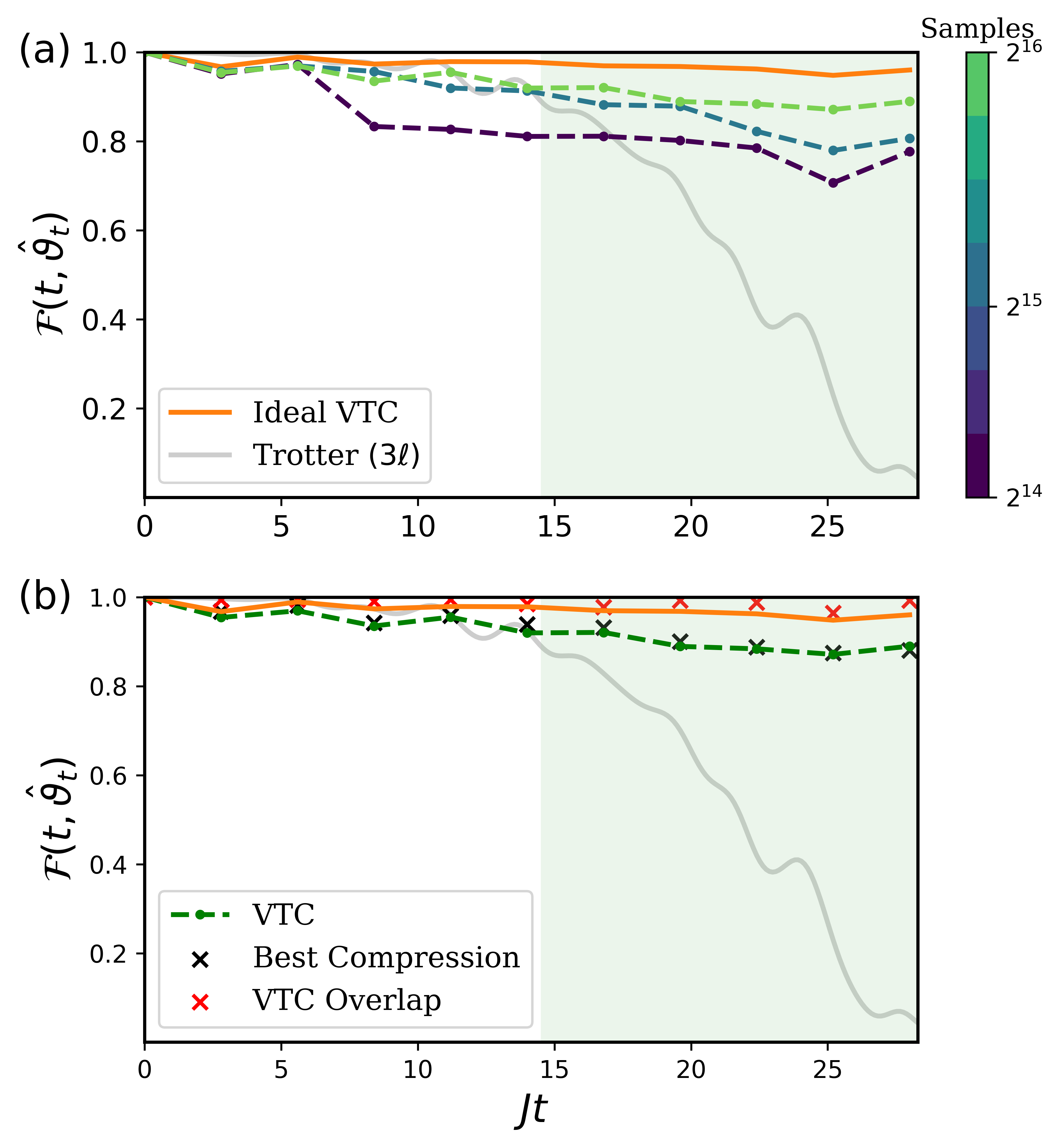}
    \caption{Optimal fidelity $\mathcal F(t,\hat{\bm \vartheta}_t)$ for VTC simulations on an ideal quantum circuit simulator. (a) Dependence of the fidelity on the number of circuit samples. Curves of different colors correspond to executions of the algorithm using different numbers of quantum measurements (samples) for each cost function evaluation, as indicated by the color bar. Results are for $M=6$, $\ell = n = 7, \tau=2.8$, and compression error threshold $\epsilon = 5\times 10^{-3}$. We use CMA-ES for the compression optimization. The direct Trotter simulations (grey) use $3 \ell = 21$ Trotter steps and are run on a statevector simulator. (b) Analysis of the fidelity data for $2^{16}$ samples in the vein of Fig.~\ref{fig:statevector_M_11}. The data points represented by black and red crosses are calculated as described in the caption of Fig.~\ref{fig:statevector_M_11}.   }
    \label{fig:shots_variation}
\end{figure}

The presence of sample noise limits the largest system size we are able to simulate using VTC to $M=6$. We do not observe any significant difference in the performance of the algorithm depending on whether we use the SWAP-test or the double time contour circuit to measure the cost function. As shown in Fig.~\ref{fig:shots_variation}(a), the fidelity for the VTC simulation is consistently larger than that for the noiseless direct Trotter simulation at large times. As expected, the VTC fidelity increases with the number of circuit samples as the sample noise is reduced. Specifically, we find a fidelity larger than $0.88$ at times $t > 25 J^{-1}$, where the fidelity using direct Trotter simulation with $3 \ell$ steps has already fallen to zero due to the accumulation of Trotter error. Note that we use a statevector simulator for the direct Trotter calculations. 

To further analyze the algorithm's performance for the maximal number ($2^{16}$) of samples, we include in Fig.~\ref{fig:shots_variation}(b) the fidelity (with respect to the exact time-evolved state) of the best possible compression during the next propagation period $\tau$. Explicitly, this is given by the overlap $|\braket{\psi(t) | U_{\text{Trot}} | \psi(\hat{\bm \vartheta}_{t-\tau})}|^2$; the data are represented as black crosses in the figure. The red crosses indicate the overlap of the best possible compression state, $U_{\text{Trot}} \ket{\psi(\hat{\bm \vartheta}_{t-\tau})}$, with the state to which the algorithm converged; explicitly, this is given by $|\braket{\psi(\hat{\bm\vartheta}_{t}) | U_{\text{Trot}} | \psi(\hat{\bm \vartheta}_{t-\tau})}|^2$. Note that this is equal to the value of the cost function $\mathcal{C}$ at the end of the optimization. An interesting feature relative to the noiseless simulation considered in Fig.~\ref{fig:statevector_M_11} is that the VTC algorithm can converge to a state whose fidelity with the exact time-evolved state exceeds that of the ``best compression" state $U_{\text{Trot}} \ket{\psi(\hat{\bm \vartheta}_{t-\tau})}$---this is visible in the last data point in Fig.~\ref{fig:shots_variation}(b). This feature is a result of the sample noise impacting the optimization during the compression step. Indeed, for the noiseless statevector simulation, the actual fidelity of the converged VTC state is upper bounded by the ``best-compression" value, as expected.

A single compression step for the parameters in Fig.~\ref{fig:shots_variation} takes up to ten hours of CPU time. The difficulty of the noisy optimization during the compression step constitutes the main bottleneck of the circuit simulations in going to larger system sizes. This bottleneck could be mitigated by parallelizing the circuit evaluations across different CPUs (or indeed QPUs), which would allow the accumulation of more samples to reduce the sample noise. Another possibility worth exploring is to replace the overlap cost function by, e.g., a reduced density matrix fidelity~\cite{Bolens_Heyl-RF-PRL-2021} defined only over a subset of the full system where local quantities of interest are to be computed.

\subsection{Noisy circuit simulator results}
\label{subsec:noisy_circuit_simulator}
We now discuss VTC simulation results on a noisy quantum circuit simulator that takes gate imperfections and finite qubit coherence times into account. To connect with the simulations on the real IBM QPU Santiago, which will be discussed below, we choose a noise model with parameters drawn from that chip, as implemented in Qiskit Aer~\cite{Qiskit}. Due to the increased noise in the simulations arising from gate errors, readout errors, and finite qubit coherence times $T_1$ and $T_2$, the numerical optimization during the compression step is even more challenging than for the ideal quantum circuit simulator. We use the non-gradient-based method CMA-ES for the classical optimization, which we find to be more reliable than gradient-based approaches.

\begin{figure}[tb]
    \centering
    \includegraphics[width=\linewidth]{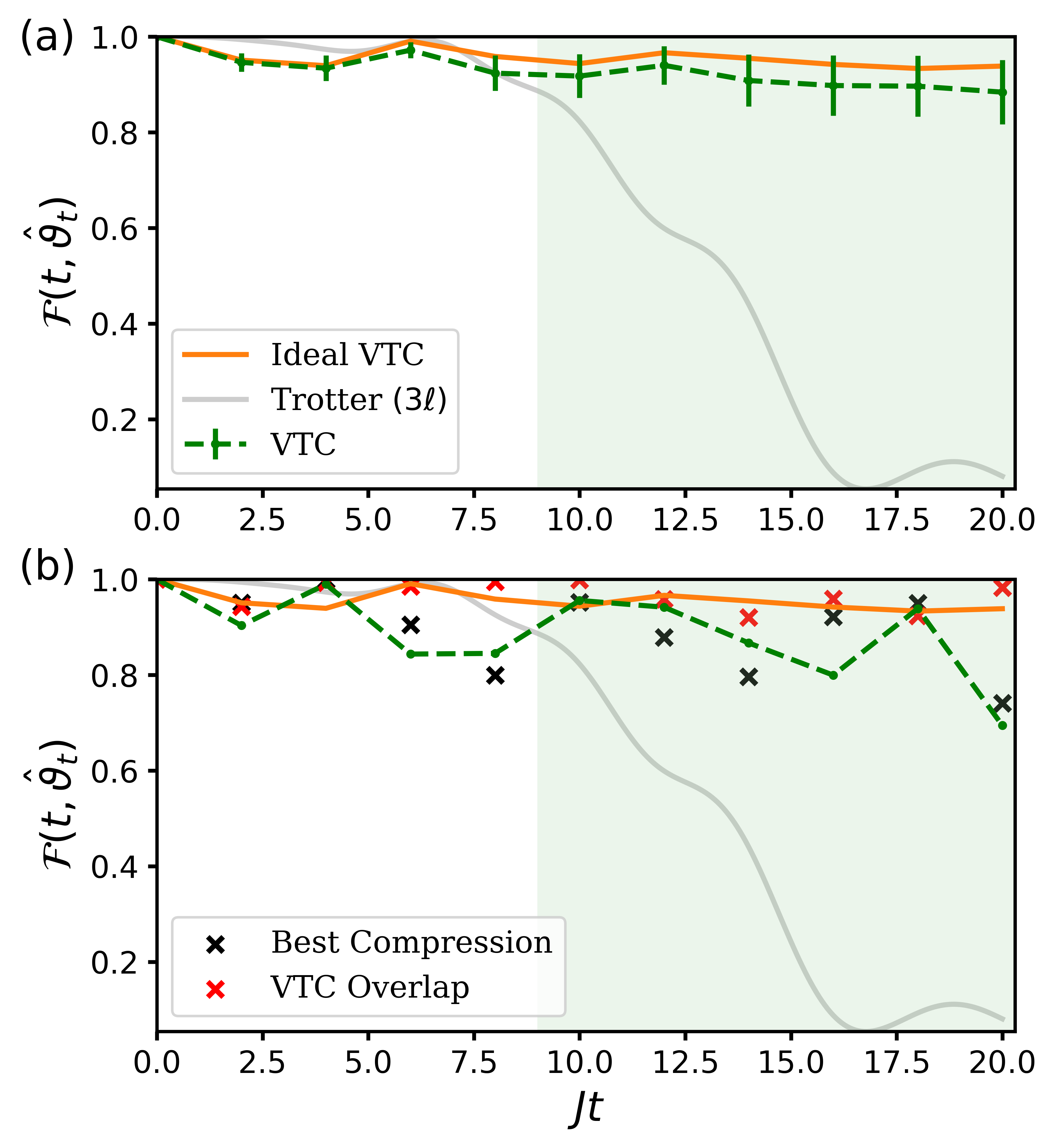}
    \caption{(a) VTC results on a noisy circuit simulator using error model parameters corresponding to the IBM Santiago backend. We use $M=3$, $\ell =n=2$, $\tau = 2$, $\epsilon=5\times 10^{-3}$, and $2^{13}$ samples. The numerical optimization is done using CMA-ES. The green dashed VTC curve shows the mean fidelity found after 50 runs of the algorithm with the standard deviation shown as error bars.
    The orange curve shows results of noiseless VTC with perfect compression. The grey curve is a noiseless Trotter simulation using $3 \ell = 6$ steps. 
    (b) VTC results on the real IBM Santiago device for the same parameters as in panel (a). Each data point on the green VTC curve represents a single run of the compression algorithm. The region highlighted in green demonstrates quantum dynamics simulations beyond the finite coherence time of the QPU. The data points represented by black and red crosses are calculated as described in the caption of Fig.~\ref{fig:statevector_M_11}.}
    \label{fig:noisy_results}
\end{figure}

To reduce the noise in the cost function, it is absolutely essential to exploit a combination of standard and specifically tailored error mitigation techniques. We employ standard readout error mitigation as built into Qiskit Ignis, and zero-noise extrapolation (ZNE) with a linear fit, which we implement using the software package Mitiq~\cite{larose2020mitiq}. We also apply a specific postselection protocol to the results: since the Heisenberg model preserves the magnetization operator $S^z=\sum^M_{i=1}Z_i$, only computational basis states with the same expectation value of $S^z$ as the initial N\'eel state are physically allowed. We thus discard any counts of computational basis states that do not fulfill the total $S^z$ conservation law after performing readout error mitigation.

In Fig.~\ref{fig:noisy_results}(a) we present VTC results for the post-quench dynamics in an $M=3$ spin chain with open boundary conditions using the noisy quantum circuit simulator. The noise parameters are drawn from the IBM QPU Santiago. We use $\ell = 2$ layers in the variational ansatz, $n=2$ Trotter steps during the VTC propagation step, and average over $8192$ circuit samples. Since the calculation of each compression point only takes about a minute, we are able to run the noisy simulation 50 times and average over the resulting VTC fidelities (green dashed line). The error bars on the green dashed line correspond to the standard deviation over these 50 runs. We observe that the noisy VTC simulations agree with the ideal VTC results (orange line) within error bars. Importantly, for times larger than $t > 9 J^{-1}$ the VTC fidelity is larger than the fidelity of a noiseless direct Trotter simulation with $3 \ell = 6$ Trotter steps. While the Trotter fidelity drops to zero around $t \approx 16 J^{-1}$, the VTC fidelity remains above $0.9$ throughout the full simulation until $t = 20 J^{-1}$.

\subsection{Results on IBM QPU Santiago}
\label{subsec:qpu_santiago}
We now demonstrate quantum dynamics simulations beyond the qubit coherence time on real IBM quantum hardware. This is achieved by running VTC simulations on the IBM QPU Santiago for a system of size $M=3$ with open boundary conditions. We use a variational ansatz with $\ell = 2$ layers, employ $n=2$ Trotter steps during the state propagation step of VTC, and average over $2^{13}$ circuit samples. The complete algorithm is executed on real quantum hardware, i.e., all circuits during the compression optimizations are being executed on the QPU. In addition to the error mitigation techniques described in Sec.~\ref{subsec:noisy_circuit_simulator}, we found it essential to apply a Pauli twirling technique, which dresses the two-qubit CNOT gates with random Pauli gates~\cite{liEfficientVariationalQuantum2017}. This is important in order to convert the noise on the real device into a stochastic form, which is necessary to justify ZNE. Without the additional Pauli twirling protocol, we observed that the ZNE error mitigation method was not reliable.

In Fig.~\ref{fig:noisy_results}(b), we present VTC simulation results on the IBM Santiago backend. These results were collected over the course of three days, since calculation of each compression point takes about 2-3 hours, which includes waiting time in the IBM execution queue. Most importantly, we find that at times $t > 9 J^{-1}$ the VTC fidelity largely exceeds the fidelity obtained from noiseless direct Trotter simulations that use the same circuit resources.
Note that the direct Trotter simulations are performed using statevector simulations and thus do not contain any of the noise that is present in the VTC simulations. 
While the direct Trotter fidelity vanishes for times larger than $t > 16 J^{-1}$, the VTC fidelity remains at a value of $0.8$ at that time. This explicitly demonstrates that even though the impact of noise on the real QPU is more severe compared to the noisy simulator, we are able to achieve quantum dynamics simulations beyond the coherence of the device using the VTC algorithm. 

In Fig.~\ref{fig:noisy_results}(b) we also analyze the data from the QPU using the methodology of Figs.~\ref{fig:statevector_M_11} and \ref{fig:shots_variation}. We find that the fidelity of the state to which the compression optimization converges can again exceed that of the best possible Trotter compression, i.e., $|\braket{\psi(t) | U_{\text{Trot}} | \psi(\hat{\bm \vartheta}_{t-\tau})}|^2$ (black crosses).
This effect, which was also apparent in the results of Fig.~\ref{fig:shots_variation}, is much more pronounced on the real QPU due to the presence of both sample and gate noise. The red crosses again represent the fidelity between the best possible compression and the state found by the algorithm. Over the $10$ compression steps, the algorithm was able to recreate the Trotter evolution of the state from the previous compression step with a mean fidelity of $0.969$.

\section{Conclusion}
\label{sec:conclusion}
We have demonstrated a simulation of the post-quench dynamics in a three-site antiferromagnetic Heisenberg chain beyond the qubit coherence time on real quantum hardware. This was achieved by compressing the Trotter time-evolved state at intermediate time steps into a variational form. The state overlap served as a cost function for the compression optimization step, which was executed on NISQ hardware without any additional overhead using a double time contour circuit. We have further benchmarked this variational Trotter compression algorithm on larger Heisenberg chains using statevector as well as noiseless and noisy circuit simulators. While most of our results were obtained for an integrable model, we also tested the ability of a variational ansatz to capture the dynamics generated by the Heisenberg Hamiltonian in the presence of an integrability-breaking term. These tests revealed that the performance of the VTC algorithm does not depend strongly on the presence or absence of integrability---indeed, an ansatz with the same number of parameters captures the wavefunction to the same accuracy in both cases.

The main goal moving forward is to increase the system sizes accessible to VTC simulations, for which several key issues need to be addressed. First, the noise level of the overlap cost function needs to be reduced in order to accelerate the classical optimiziation that is performed during the compression step. In addition to reducing device errors by hardware improvements, one can devise several error mitigation strategies~\cite{endoHybridQuantumClassicalAlgorithms2021} such as probabilistic error cancellation~\cite{temmeErrorMitigationShortDepth2017}, virtual distillation~\cite{hugginsVirtualDistillationQuantum2021,koczorExponentialErrorSuppression2021}, Clifford data regression~\cite{czarnikErrorMitigationClifford2021} or a combination thereof~\cite{caiMultiexponentialErrorExtrapolation2021, mariExtendingQuantumProbabilistic2021, loweUnifiedApproachDatadriven2021, bultriniUnifyingBenchmarkingStateoftheart2021,piveteauQuasiprobabilityDecompositionsReduced2021}. Sample noise can be reduced by increasing the number of circuit shots beyond the current limit of $2^{13}$ on the IBM backend we used. The fidelity of the Trotter time evolution during the propagation step can be increased by using pulse level control of the gates, which can often lead to shorter gate times. Finally, other future directions would be to explore alternative variational ansaetze such as hardware efficient ones, and to employ other classical optimizers for noisy cost functions.

\acknowledgments
The authors acknowledge valuable discussions with M. Sohaib Alam, Stephan Rachel and Yong-Xin Yao.
This material is based upon work supported by the National Science Foundation under Grant No.~2038010.


%

\end{document}